# EXSCLAIM! – An automated pipeline for the construction of labeled materials imaging datasets from literature

Eric Schwenker[1,2], Weixin Jiang[1,3], Trevor Spreadbury[1,3], Nicola Ferrier[4], Oliver Cossairt[3], Maria K. Y. Chan[1]

1. Center for Nanoscale Materials, Argonne National Laboratory, USA
2. Department of Materials Science and Engineering, Northwestern University, USA
3. Department of Computer Science, Northwestern University, USA
4. Mathematics and Computer Science, Argonne National Laboratory, USA

corresponding author: Maria Chan (mchan@anl.gov)

**Abstract**

Due to recent improvements in image resolution and acquisition speed, materials microscopy is experiencing an explosion of published imaging data. The standard publication format, while sufficient for traditional data ingestion scenarios where a select number of images can be critically examined and curated manually, is not conducive to large-scale data aggregation or analysis, hindering data sharing and reuse. Most images in publications are presented as components of a larger figure with their explicit context buried in the main body or caption text, so even if aggregated, collections of images with weak or no digitized contextual labels have limited value. To solve the problem of curating labeled microscopy data from literature, this work introduces the EXSCLAIM! Python toolkit for the automatic **EX**traction, **S**eparation, and **C**aption-based natural **L**anguage **A**nnotation of **IM**ages from scientific literature. We highlight the methodology behind the construction of EXSCLAIM! and demonstrate its ability to extract and label open-source scientific images at high volume.

**Introduction**

Journal articles have long been the standard for communicating important advances in scientific understanding. As sophisticated measurement and visualization tools render scientific communication more intricate and diverse, the visual presentation of scientific results as *figures* in these articles is noticeably more complex – especially within journals considered high-impact [1]. With this complexity, which is often a byproduct of the "compound" layout of the figures (*i.e.* figures containing multiple embedded images, graphs, and illustrations, *etc.*), the meaning of each individual image as a standalone entity is not always apparent. The result is that individual images are not only unsearchable, but the effort required to extract them into a machine-readable format is significant. This plays a major factor in the relative scarcity of general materials characterization images in the development and testing of new algorithms in deep learning (DL), an emerging field characterized by the use of deep neural networks – hierarchical, multi-layered structures of processing elements – to learn representations of input data (often images) that reveal important characteristics about its content or overall appearance. The current surge of interest in DL stems from recent success in applications such as facial recognition [2], self-driving cars [3], and complex game playing [4]. Much of this success is the byproduct of having large labeled datasets available for training [5], and in order for current materials imaging classification and recognition tasks [6]–[14] to reap the benefits afforded by DL pipelines, having access to substantial labeled data is crucial. Fortunately, the incentive to publish is nearly ubiquitous across all scientific disciplines, and with a mechanism for automatic dataset construction that includes both separating out individual images from compound figures, as well as providing concise annotations describing key aspects or classification of the image content, much more of the scientific imaging data in literature could be utilized for training and developing DL models.

The effort to automate the construction and labeling of datasets from general web data has garnered broad attention from the computer vision, language technologies, and even chemistry/materials informatics communities [15]–[18]. From the chemistry and materials informatics perspective, most of the focus has been on the development of text-mining tools adapted for "chemistry-aware" natural language processing (NLP), and have been used to create datasets of material properties and synthesis parameters from journal article text [19]–[24]. For imaging datasets, standard computer vision approaches take the top images retrieved from a keyword query of an image search engine (i.e. Google, Flikr, *etc*.) and train classifiers to further populate datasets based on a keyword [18]. Unfortunately, this approach is problematic for scientific figures because of their compound layout. Current works that address this layout problem (what we refer to as "figure separation") rely on hand-crafted rule based approaches [25]–[29], or adapt neural-networks to interpret figure separation as an object detection problem [30], [31]. While handcrafted techniques generally work well for sharp image boundaries, and neural network approaches capture irregular image arrangements, neither of these methods are designed such that explicit references to the caption text are considered as part of the separation. This is problematic for both figure separation and labeling because when the two are not consistent with each other (*i.e.* more/fewer images than subfigure labels) the intended description for an image is not always clear. In these instances, (NLP) with extra-grammatical constructs [17], can be extended to properly summarize pertinent scientific results consistent with the subdivision of the image. Despite the individual successes of a few previous studies to advance figure separation, and labeling related to automatic dataset creation [25], [26], [32], tools that do this in a seamless end-to-end (query-to-dataset) fashion, capable of extracting and labeling images from journal articles based on the specific user requirements, are currently lacking.

In this work, we present a tool for the automatic **EX**traction, **S**eparation, and **C**aption-based natural **L**anguage **A**nnotation of **Im**ages (EXSCLAIM!) for scientific figures and demonstrate its effectiveness in constructing a self-labeled electron microscopy dataset of nanostructure images. The EXSCLAIM! tool is developed around materials microscopy images, but the approach is applicable to other scientific domains that produce high-volumes of publications with image-based data as well as graphs and illustrations.

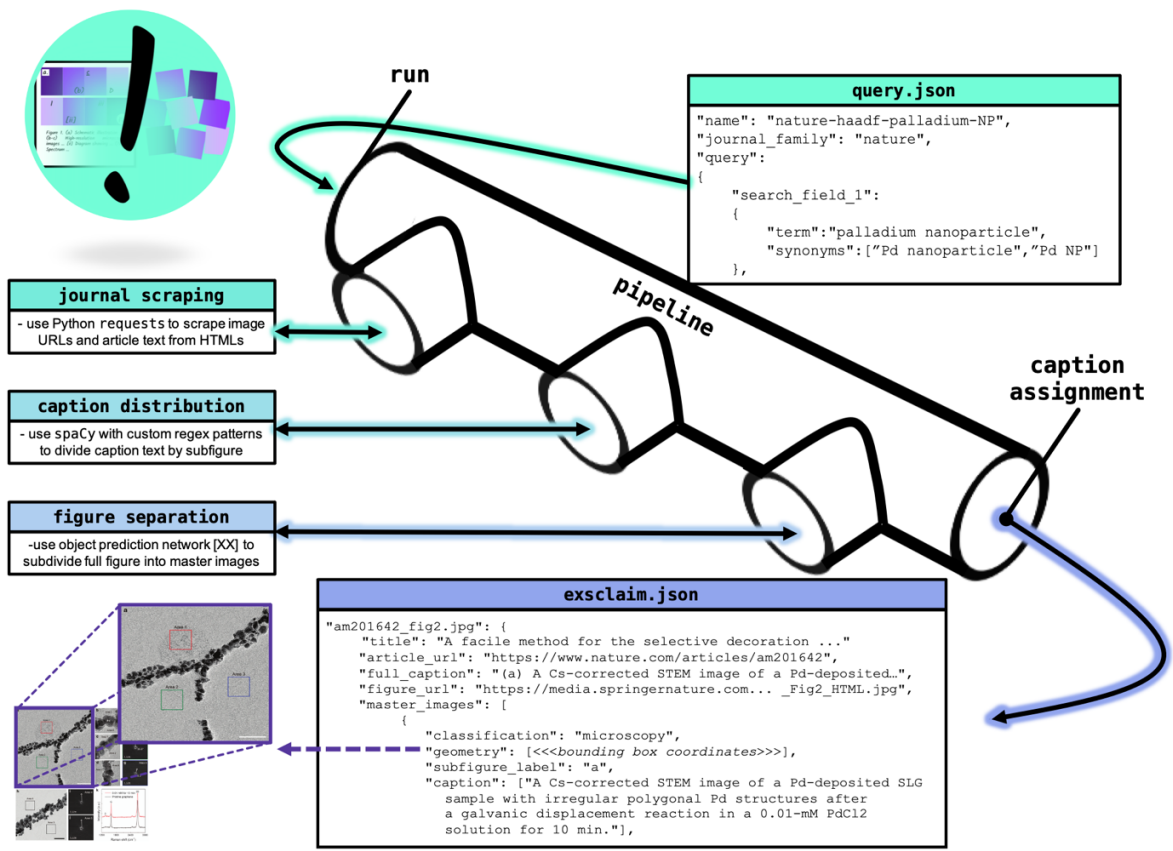

**Figure 1: An Overview of the EXSCLAIM! Pipeline.** The schematic highlights the path from a user-defined *query* to the *exsclaim* output data structure (both JSON objects). After the query is submitted to the pipeline by calling the *run* method, the *journal scraper* extracts figure/caption pairs from articles in the specified journal family, the *caption distributor* divides the caption text into segments that are consistent with the images in the figure, and the *figure separator* computes bounding boxes that separate and classify the individual images from the full figure. The *caption assignment* method is effectively the "self-labeling", as it takes all of the caption segments, reduces them to individual keywords if possible, and pairs them with their corresponding image within the full figure. The final output data structure, *i.e.* the *exsclaim* output file, contains all the descriptions and references necessary to construct the full labeled imaging dataset.

## Methods

*Design Overview.* The main goal of the EXSCLAIM! toolkit is to provide researchers with a collection of modules that, when executed as a sequential pipeline (1) enable comprehensive keyword searches for scientific figures within open-source journal articles, and (2) facilitate the extraction and pairing of images from within figures, to the appropriate descriptive keywords and phrases from caption text. To begin, users populate an input file (JSON object) with parameters that define the search (*i.e.* keywords, synonyms, how to order search results, journal family to search, *etc.*), as well as the number of total articles to consider in the process. This input file, referred to as the *query*, is the basic initialization structure that is *run* in the EXSCLAIM! pipeline. A *run* of the EXSCLAIM! pipeline involves sequential execution of the *journal scraper, caption distributor,* and *figure separator* modules before a final call to the *caption assignment* module which is responsible for the final pairing of the distributed caption text to the extracted figures. Figure 1 provides an overview of the pipeline, highlighting the structure of both the *query* and final *exsclaim* output file (another JSON object, summarized for better readability), as well as the role that the extraction modules serve in transforming the data. In Fig. 1, the *query* for palladium nanoparticles is run through the pipeline, and each extraction module incrementally populates the *exsclaim* output file, which provides a final record of all the relevant figures with bounding box and label information for the associated images. This *exsclaim* output file provides all the necessary file identifiers and pointers needed to construct a high-quality self-labeled imaging dataset based on the details of the overall search as defined by the *query*.

*Journal Scraper.* This module performs the first extraction step in the pipeline. It is responsible for retrieving figure/caption pairs from articles that fit the parameters of the search defined in the *query*, and uses the python *requests* library to handle all of the HTTP requests. First, article URLs are extracted from a collection of individual searches formed from all possible combinations of search field keywords and associated synonyms. With an ordered list of open-source article URLs, the scraper contains a method that sends further GET requests to retrieve each article and its figure/caption pairs. Finally, the *exsclaim* file is populated with all the general article information returned during the extraction steps, including the full caption data for each figure, and URLs to the articles and their associated figures – providing basic provenance to the data workflow. (Full article text can also be included if desired.) The *journal scraper* has the functionality to parse both open-source Nature family journals as well as journals that are part of the American Chemical Society (ACS) umbrella. All of the figure and caption extraction is performed directly from the HTML version of the article and does not require PDF downloads.

*Caption Distributor.* With all of the figures and caption information recorded, the next step is to distribute subsequences of the full caption text to the respective subfigure that they reference. This first involves sentence tokenization with Parts-of-Speech (POS) tagging. POS tagging deconstructs strings of sentence text into small units (tokens), which are given a tag that describes their part-of-speech in context of the sentence. For this, the robust natural language processing (NLP) tokenization tools from the spaCy library [33] are extended to properly assign "subfigure identifier" tags, to patterns that indicate the presence of a subfigure description. With this custom tagging, the "(a)" in a phrase such as *"(a) Nanoparticles deposited on ...",* is properly interpreted as the subfigure identifier (denoted by 'CAP') instead of a determiner surrounded by parenthesis. From there, a regular expression style of *pattern matching* is performed on the list of the custom POS tags with a dictionary of reference sequences collectively representing a "standard syntax" for typical subfigure image descriptions in caption text. For example, Table 1 shows the full caption text divided according to POS tags and the pattern, designated as "Pattern #1", is used for the first search. It is looking for a collection of POS tags that begin with a subfigure identifier, proceeded by a noun chuck (a noun phrase), a preposition, and then any number of POS tags are allowed until a full stop. This specific sequence would successfully extract a caption description for image (a) as *"TEM images of 1.93 wt% Ru–WSe2"* from the full caption text. From here, words in the caption text that are search terms in the *query* are designated explicitly as "keywords" for the image.

*Figure Separator.* The final component of the EXSCLAIM! pipeline involves separating the extracted figures into "master images" according to what we establish as the Master-Dependent-Inset (MDI) modeling paradigm. In the MDI model, the master image is defined relative to a subfigure label ("(a)", "(b)", etc.). The subfigure label is the functional element bridging the visual image content to the text describing it. All visual components (i.e. all individual images, drawing, clarifying annotations) belonging to the complete entity referenced by the subfigure label, regardless of the shape it forms when all entities are considered together, are collectively referred to as the "master image". In addition to defining the master images in context of the full figure, the *figure separator* module both classifies the image according to the nature of the image content, *i.e.* microscopy, diffraction, graph, illustration image *etc.,* and also

**Table 1:** Custom sentence-level regular expressions applied to a POS-tagged caption with the assigned text strings.

| |
|---|
| **Full Caption Text (custom POS tagging)**<br>from: Y. Zhao, G. Mao, C. Huang, P. Cai, G. Cheng, W. Luo, *Inorg. Chem. Front.* 2019, *6*, 138<br>(a) and (b) TEM images of 1.93 wt% Ru-WSe$_2$. (c) HRTEM image of 1.93 wt% Ru-WSe$_2$. (d) and (e) The enlarged area denoted in (c) corresponds to the HRTEM images of WSe$_2$. (f) HAADF-STEM image of 1.93 wt% Ru-WSe$_2$. (g-i) The EDS mapping of Ru, W, and Se, respectively. |
| **Caption Text Selections (with custom POS tagging)**<br>('(a) and (b)', 'CAP'),('TEM images', 'NC'),('of', 'IN'),('1.93 wt% Ru-WSe2', 'NC'),('.', '.') … ('(d) and (e)', 'CAP'),('The enlarged area', 'NC'),('denoted', 'IR'),('in', 'IN'),('(c)', 'CAP'),('corresponds', 'IR'),('to', 'IN'),('the HRTEM images', 'NC'),('of', 'IN'),('WSe2', 'NC'),('.', '.')… |
| **Pattern #1:**<br>("CAP", "!", "NC", "IN", "*", ".")<br>- i.e. start with a caption delimiter and do not record any text until a noun chunk is found that contains a proposition immediately after. From there include all text until a full-stop is detected |
| **Text assigned to (a) and (b) using Pattern #1**<br>(a)     TEM images of 1.93 wt% Ru-WSe2.<br>(b)     TEM images of 1.93 wt% Ru-WSe2. |

extracts any scaling information that is present in the form of a scale bar on images within the figure. Figure 2 provides a detailed view of the MDI model applied to a standard figure. Both insets and scaling information are shown in subfigure "(a)", and subfigure "(b)" is useful for demonstrating the need for a "master image" classification, as it is clear that the subfigure label is referencing more than one distinct image. Moreover, when a master image contains multiple dependent images, it is classified as a parent. The detection and classification of master images is the primary task of the *figure separator* module, which follows a two-stage framework outlined in Jiang *et al* [34]. The first stage is subfigure label detection, where the goal is to detect all of the subfigure labels presented in the compound figure. This is achieved using a combination of object localization (YOLOv3-style object detector [35]) and object recognition (ResNet-152 [36]) neural networks. The second stage starts with encoding the global subfigure label layout information, which is extracted from the detected subfigure labels, into a binary mask. Then this binary mask, which provides visual anchors for the positions of the subfigure labels, is concatenated with the standard RGB input channels and fed into the master image detection module. Taken together, these neural networks function to locate master images within the figure while preserving the association between master images and their respective subfigure labels, and further classify each master image based on the appearance of its content.

After the master images are detected, localized, and classified, the final step is to extract any scaling information that exists. Scale detection also proceeds through a two-stage network. First an object detection neural network (Faster R-CNN [37]) is used to detect the bounding boxes of scale bar labels and scale bar lines that exist in a given figure. Next, the detected scale bar labels are fed into a Convolutional Recurrent Neural Network (CRNN [38]) in order to perform text recognition, making the scale bar label text machine readable. The result of the CRNN is processed by a rule-based search to ensure the output is a valid scale bar label (i.e. a number followed by a unit). Multiple scale bar lines and scale bar labels in a single figure are matched by assigning the scale bar labels to scale bar lines greedily based on the distance between the center of their respective bounding boxes. Each matched scale bar-scale bar label pair is assigned to the subfigure in which it is contained. Using the length in pixels of the scale bar line and the subfigure and the scale bar label text, the real space size of the subfigure can be determined.

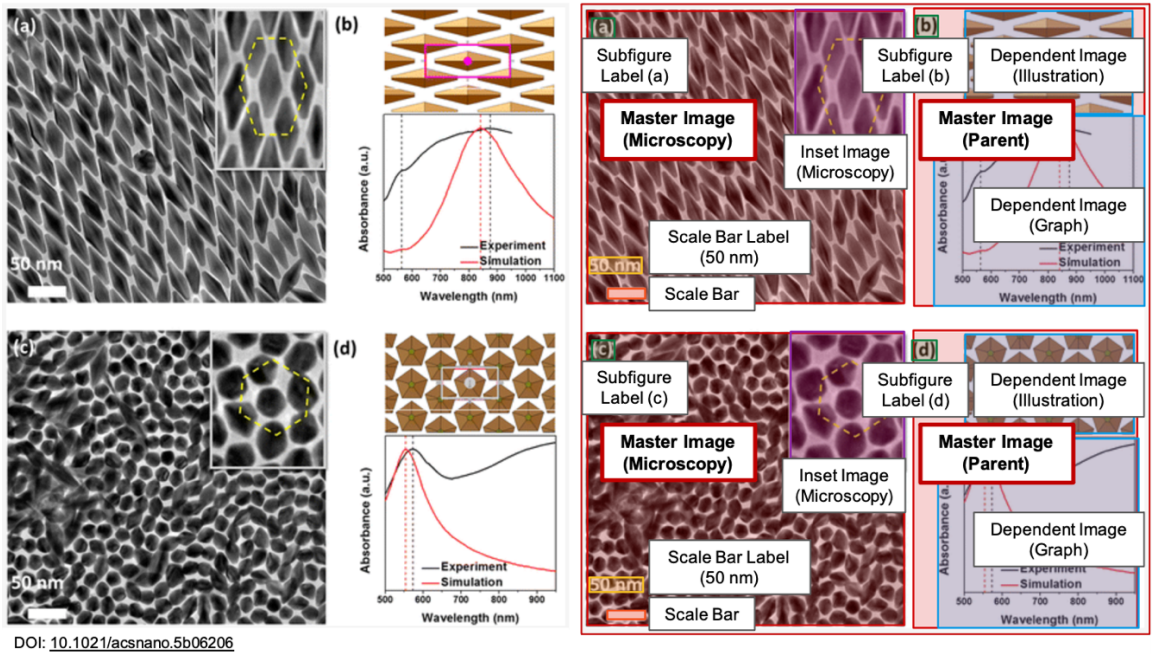

**Figure 2: Master-Dependent-Inset (MDI) Model**. The panel on the left shows the initial figure before annotation, and the panel on the right is a copy that has been properly annotated according to the Master-Dependent-Inset (MDI) model. The presence of a subfigure label is necessary for all images given a "master image" designation. Notice the master image corresponding to subfigure label "b" contains two dependent images (illustration and graph). Since the master image corresponding to subfigure label "b" governs more than one distinct image (i.e. the caption text associated with "b" will refer to all the images in the series), this master image is classified as a parent. Additional image features such as the scale bar and scale bar label (when present) are also identified.

*Crowdsourcing Figure Annotation with Mechanical Turk.* Ultimately, DL models are needed for detecting the locations of the master images and classifying these images by type. In order to achieve sufficient accuracy, these models must be trained on images that are deemed as proper references, or have been verified as representing the correct way to locate and classify master images (*i.e.* ground truth). Because the demands of this task are unique in that figure separation does not fall explicitly within the canon of standard computer vision training tasks, we needed an approach to scale the annotation effort to ensure the best accuracy on the figure separation task. For this, we used the crowdsourcing platform from Amazon called *Mechanical Turk* (MTurk). Though proper interpretation of a scientific image often requires an expert to understand the nuances of the image content, identifying where the master images are located, as well as their proper classification, can be formulated so that those without a rigorous science background can annotate the images with only a very modest amount of instruction. As such, we designed a custom figure annotation GUI (snapshots of the GUI are included in the S.I.) within the MTurk platform, and asked workers to draw bounding boxes around each master image in the dataset, and then classify them. This allowed us to quickly create a dataset of > 3000 MDI annotated figures (~18,000 separate images). The basis for training the current version of the *figure separator* involves augmentation of a random sampling of 2000 of the annotated images from MTurk and is described in more detail in Jiang *et al* [34].

# Results

The extraction modules, outlined individually in the Methods section, incrementally transform the query into a final *exsclaim* output structure, which contains all of the information necessary to create a dataset annotated from caption descriptions in the literature. There are several components of this pipeline that must be considered in evaluating overall performance. Here, we (1) validate classification accuracy for the *figure separator* tool using precision and recall metrics obtained on a reference data set, (2) examine the various scenarios for how caption text is assigned, quantifying accuracy for the case where a single keyword is used to describe the image, and finally (3) provide suggestions for how to create new labels or general topics to associate with images that are left un- or under-annotated. In total, the results emphasize the attention placed on accuracy and extensibility of the EXSCLAIM! pipeline design.

### *Validation of the figure separator tool on MTurk Dataset*

In order to validate the classification and bounding box prediction accuracy of the *figure separator* tool, 784 figures (3555 separated images) from the crowdsourced MTurk dataset – withheld during training – were used as part of the validation set. The validation dataset is available from the Materials Data Facility [39], [40] (DOI: 10.18126/a6jr-yfoq). The results are shown in Figure 3. Images were scraped from Nature publishing sources (the code for the scrapers is easily extendable to other journal sources), and positive predictive value (precision), which is loosely the correctness of the positive classification, is always prioritized over a measure of completeness, such as recall. Confusion matrices summarizing important aspects of the classification performance are given in both a "no threshold" (Figure 3a) and "high threshold" (Figure 3b) condition. In the "no threshold" condition, the most likely classification in the final output of the neural network is accepted, regardless of its magnitude. In the "high threshold" condition, only classifications with values of magnitude greater than or equal to 0.99 are accepted. In this context, the threshold is a proxy for classification confidence in the *figure separator*. Both microscopy and graph classification with "no threshold" and "high threshold" specification are favorable from a precision perspective with all scores in excess of 0.80 – and furthermore, the "high-threshold" precision for graphs is ~0.98.

One of the primary use cases for the EXSCLAIM! toolkit is the construction of self-labeled datasets, and in the assumption of data abundance (*i.e.* the opportunity cost of passing up on an image is low), the low recall values from false negatives (particularly diffraction images identified as microscopy images) are not as detrimental to the integrity of the set as images that are incorrectly identified as an image type they are not (i.e. false positives). Figure 3c highlights some of the interesting false positive trends where a sizable population of diffraction, illustration, and parent classes are being incorrectly assigned as microscopy images. In the case of the diffraction example (left), the diffraction patterns show periodicity reminiscent of atomic-resolution microscopy images, so the microscopy assignment seems logical. The false positive for an illustration (middle) maintains some features commonly associated with a microscopy image, such as a darker background, and even the coloring of the graphene sheet resembles a texture present in microscopy images, but globally is clearly not a microscopy image. Finally, the parent on the far right (Figure 3c) actually does contain a microscopy image, but because the image below it does not contain a subfigure label and is "semantically" tied to the microscopy image in "a", the most appropriate classification for this image is "parent".

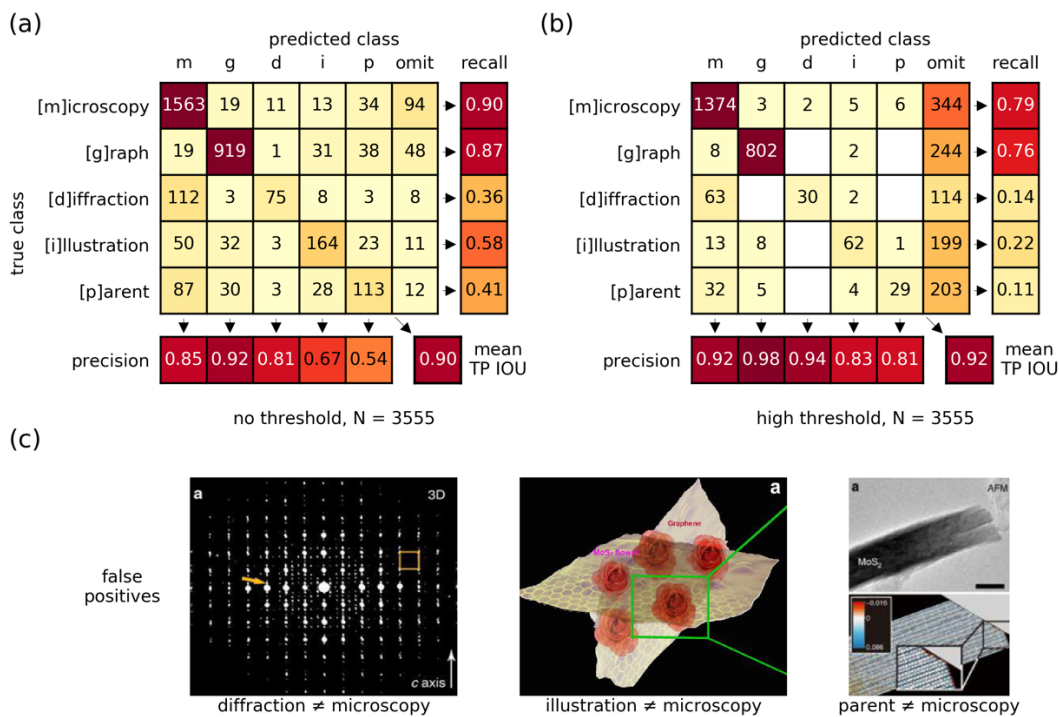

**Figure 3: Precision and Recall for Image Classification.** The confusion matrices highlight the nature of the mistakes made in each classification scenario at two confidence thresholds (a) no threshold, and (b) high threshold for N=3555 images. In both cases, the precision scores are adequate, particularly in the case of the microscopy, graph, and diffraction images. Recall suffers across the board as correctness is emphasized over completeness in the design of the pipeline. Images in (c) highlight some of the more easily rationalized examples of false positive microscopy classifications. | DOIs for articles containing the example images (left to right): 10.1038/ncomms14925, 10.1038/srep08722, 10.1038/ncomms4631.

### *Sample Query: Electron Microscopy Images of Nanostructures*

We illustrate the utility of EXSCLAIM! for labeling materials imaging datasets with an example of electron microscopy images of nanostructures. Open source Nature articles were collected from a "Sort By Relevance" search related to the collection of queries formed from the following lists of keywords: ("electron microscope", "electron microscopy"), and ("nanoparticle", "nanosheet", "nanoflake", "nanorod", "nanotube", "nanoplate", "nanocrystal", "nanowire", "nanosphere", "nanocapsule", "nanofiber"). This specific query mimics a wildcard-style search for nanostructures imaged in an electron microscopy modality and returned a total of 13,450 open-source articles with 83,504 figure-caption pairs. For the purpose of quantifying overall retrieval performance on microscopy images, which involves both an assessment of image classification and keyword labeling accuracy, we restrict the following quantitative measurements to articles in the top 10% of the relevancy ranked list, which is a reasonable simplification because the labels defined by the *query* ("nanoparticle, "nanowire", etc.) depends on the presence of the keyword in the caption, and the median keyword frequency decays exponentially across article rank (refer to the S.I.). This collection of articles has a yield of 29,096 separate images. The full dataset returned from this *query* of nanostructure images is available from the Materials Data Facility [39], [40] (DOI: 10.18126/v7bl-lj1n).

Graphs and microscopy images are among the most popular image types for this specific *query*. The high prevalence of microscopy images is expected, as a result of including microscopy-relevant keywords as a separate word family, whereas the high frequency of graphs is most likely a result of how authors choose to format scientific results. Figure 4a highlights the distribution of predicted image types in the top 10% of retrieved articles, and indicates the threshold used to assign the classification with further color coded divisions of each bar. These thresholds ("no threshold" and "high threshold") again act as proxy for classification confidence and are consistent with the definitions given in the evaluation of the figure separator on the MTurk dataset. Within both graphs and microscopy images, ~75% of the total population receives a high-level of confidence associated with the prediction of its image type (*i.e.* high threshold condition), which is likely a consequence of having a more precisely defined image type. This distribution of predicted image types in Fig 4a is useful for quantifying the approximate number or percentage of high-confidence extractions (image classification) that one could expect to obtain for a given query. Moreover, taken with the results on the MTurk images from Fig 3, it is likely that a large fraction of the high-confidence classifications, particularly in the case of microscopy images and graphs, are actual instances of microscopy images or graphs because of the high precision scores.

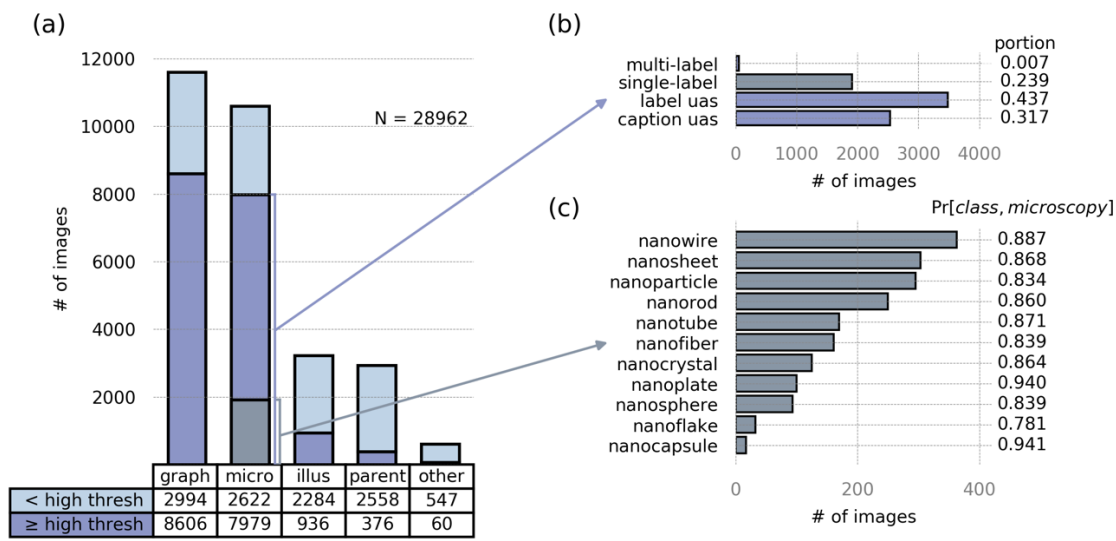

**Figure 4: Distribution of Image Types and Keyword Labeling Accuracy.** The *query* in this example is used to extract electron microscopy images of general nanostructures from Nature journals. The bar plot in (a) shows the distribution of image types extracted at two different thresholds. The bar plots in (b) and (c) further subdivide the population of high confidence microscopy images. In (b), the distribution of label types are recorded. In this context, single and multi-label refer the existence of a keyword label. Label unassigned (uas) means that caption text has been distributed to the image, but no keyword label from the *query* exists. Caption unassigned (uas) refers to a scenario where the *caption distributor* was not able to confidently distribute a proper substring to caption text. The bar plot in (c) represents the distribution of labels in the top 10% of retrieved microscopy images and provides a estimate of the joint probability that an image classified as microscopy and given the corresponding subsequent label is a true microscopy image represented by the given label.

To give classification confidence some further meaning in the context of the construction of self-labeled microscopy image dataset, it is important to start examining both the frequency and quality of the processed caption text during *caption assignment*, because it is this text that is used to ultimately describe the image content (*i.e.* it is the difference between constructing a generic dataset of microscopy images vs. a highly specific dataset of atomic resolution microscopy images of Ag nanoparticles). First, we examine the frequency at which the processed text is assigned to images and look further at the nature of the text extracted (*i.e.* is it a single keyword? multiple keywords? sentence or phrase?). Figure 4b identifies four categories of possible image labeling. The most common case, "label uas" (label unassigned) represents a population of the separated images that have received a portion of the caption text as a description, but do not contain any of the keywords from the *query*. The "caption uas" (caption unassigned) category, contains separated images that have not received a portion of the caption text at all. These captions that should be assigned typically have sentence structure complexities, such as multiple compound subjects, or long intervening phrases, etc., that make caption distribution somewhat ambiguous. And while the percentage of "caption uas" images is somewhat high, this is to some degree intentional, as the data abundance assumption and "no information is better than bad information" design principles underscore many of the extraction steps in the pipeline for the purposes of keeping the data clean. Also, the current regular expression approach taken in caption distribution provides users with the ability to extend or fine tune the system to better suit individual use cases. The final conditions, identified in Fig 4b, are the single and multi-label conditions, which represent images containing assigned caption text with explicit reference to one or more of the *query* keywords.

The single-label condition is further broken down in Fig 4c, and its relation to the initial full set of extracted microscopy images is emphasized with the gray coloring in Fig 4a (~24% of the images in the high-threshold group have a single keyword label). The bar chart shows the distribution of images assigned to each keyword, as well as a measure of the joint probability of both the predicted microscopy classification and keyword as being correctly identified. The average joint probability of the positive microscopy/keyword prediction computed across all classes is ~87%, which means that of the 1909 single-label microscopy images, 1660 are true microscopy images with appropriate keyword labels. Across the entirety of the extracted images (beyond the results for the top 10% recorded here), there are approximately 4300 microscopy images with a single keyword explicitly related to the query. With the ~87% joint probability of the positive microscopy/keyword prediction, we would expect a full dataset for this *query* to contain approximately 3725 high-confidence microscopy images with the appropriate keyword label.

Current caption distribution prioritizes identification of appropriate keywords over language correctness, so only keyword accuracy is factored into the ground truth comparisons in Figure 4. There are, however, many examples where both the keywords are successfully extracted, and the distributed caption is grammatically sufficient, and this even extends to scenarios where the caption text assignment contains non-contiguous segments. For example, in Figure 5a, the distributed caption uses all the text from the subfigure identifier "(b)" to the period signifying the end of the sentence, however Figure 5b-c show that the current method is capable of parsing the sentence at a higher structural level and pick out the structurally closest sequence of text, even when it is not linearly the closest. For example, in the Figure 5c, the subject of the sentence "TEM images" is not matched to the linearly closest descriptions (adjectives) or even closest preposition, but in this case to the preposition at the end of the sentence, which structurally makes sense and is the proper division of this caption text.

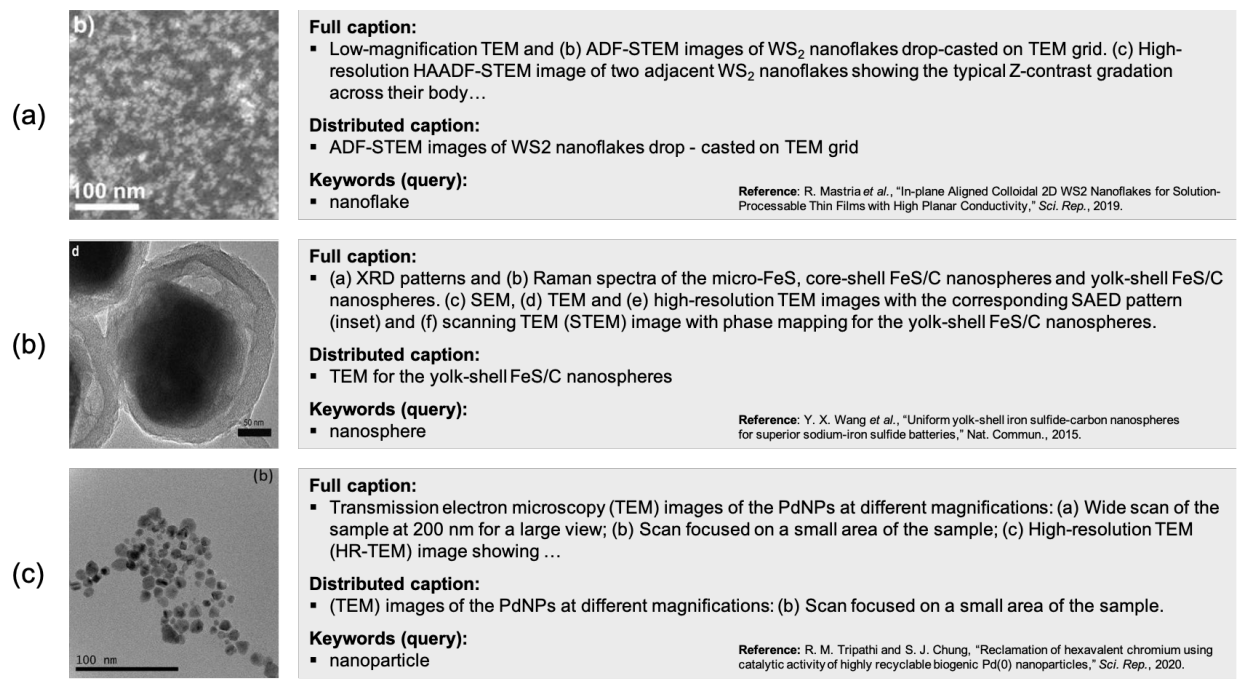

**Figure 5: Examples of Images Extracted and Labeled with EXSCLAIM!** The extracted images contain both a keyword label from a word family associated with the *query*, as well as grammatically sufficient sequence of distributed caption text. Some caption distribution examples are simple, as in (a) where all the words distributed are linearly connected, however, the current *caption distributor* extraction class is also designed to adequately capture more complex structural dependence relationships (b-c) where the subject is separated from the text completing the full consistent description.

### *Extraction of Scaling Information from Images*

Magnification of objects is fundamental to microscopic imaging. With suitable estimation of object magnification, achieved by recognizing and interpreting the scale bar length and accompanying text (scale bar-scale bar label) in the separated subfigures images, researchers can begin quantifying image content based on size. In addition, searches through resulting databases can be refined to a particular size range, giving users greater control over results. Further modeling can combine scale and caption information to associate keyword terms with dimensions. To quantify the accuracy of the scale detection step, 440 figures containing 920 scale bar-scale bar label pairs from the MTurk dataset that were withheld during training were used. The predicted pixel length of the scale bar line differed from the ground truth value by a mean absolute error of 5.4%. This level of error is similar to that of humans performing the task and negligible when determining the approximate scale of the image. Scale bar label recognition overall is 82% accurate (both number and unit) (See Figure S3 in the SI) when the confidence threshold is 0.2. This is a reasonable accuracy given the variance in the quality of the scale bar text itself. It is not uncommon for authors to leave the default scale bar text untouched when a microscopy image is included as part of a compound figure – this is problematic because the native text size is often too small for high quality visualization, and there are many instances where the color of the text is similar to the image content it sits on top of (i.e. recognition suffers from very low contrast). Refer to some of the examples included in the SI which support these general observations. The SI also includes confusion matrices to summarize the prediction results breaking each label down by number and unit.

*Self-labeling with NLP tools*

We have demonstrated the effectiveness of the EXSCLAIM! tool in situations where keywords are extracted from the caption text, and we even show that in some situations when the structure of the caption is complex, caption assignment is still capable of extracting sentences and phrases in a way that preserves the intended meaning. However, currently, keywords are defined explicitly by the *query*, so any other related topics that are prominent in the returned set, but not explicitly specified within the *query*, are ignored. The advantage of treating the problem of dataset construction in this manner is that it ensures a high degree of relevance in the images that are assigned a keyword label, but it significantly limits the scope of the self-labeling task in general (i.e. Fig 4b. reveals that ~43% of the images in the nanostructure example contained distributed caption text but were without explicit keyword labels). To this end, we explore how modern natural language processing (NLP) tools can be leveraged to transform the phrases or sentences of the distributed caption text into a series of relevant, hierarchical labels for each image they are referencing. In particular, the outlined approach involves the use of two popular techniques in NLP: word embeddings and topic modeling from documents.

*Word Embeddings.* The goal of word embeddings is to create vector space representations for individual words in such a way that similar words are located close to one another in vector space. EXSCLAIM! leverages the popular unsupervised Word2Vec technique [41] to learn high quality word vectors for the images returned in the nanostructure query, using the 26,683 abstract and introductory paragraphs (*i.e.* all text before methods are described) from the source articles. Constraining the language to the topics present in the searched articles is appropriate for the goal of image self-labeling. For a more comprehensive embedding of the materials science literature with an information discovery focus, refer to Ref. [24]. To demonstrate how Word2Vec can be used for word associations in context of the abstracts and introduction texts, Figure 6a highlights simple word lookup examples. Without being explicitly associated, "nanoparticles" and "nanowire" are placed in close proximity to their abbreviations, "nps" and "nw", respectively. Additionally, 3D "nanoparticles" are closely associated with another 3D nanostructure, such as a "nanocrystal", and a "nanowire" is placed in close proximity to a similar 2D "nanorod". The "angstrom" unit of length is placed closely to "nanometer" and "micrometer", and interestingly, in this case, scale is even preserved in the ordering (i.e. angstrom is closer in scale to a nm than micrometer). These sorts of quantitative relationships are not uncommon and are described in further detail in the original paper [41]. The "tem" and "cnt" lookups provide further demonstrative evidence of the Word2Vec's effectiveness in creating some notion of proper word associations and context.

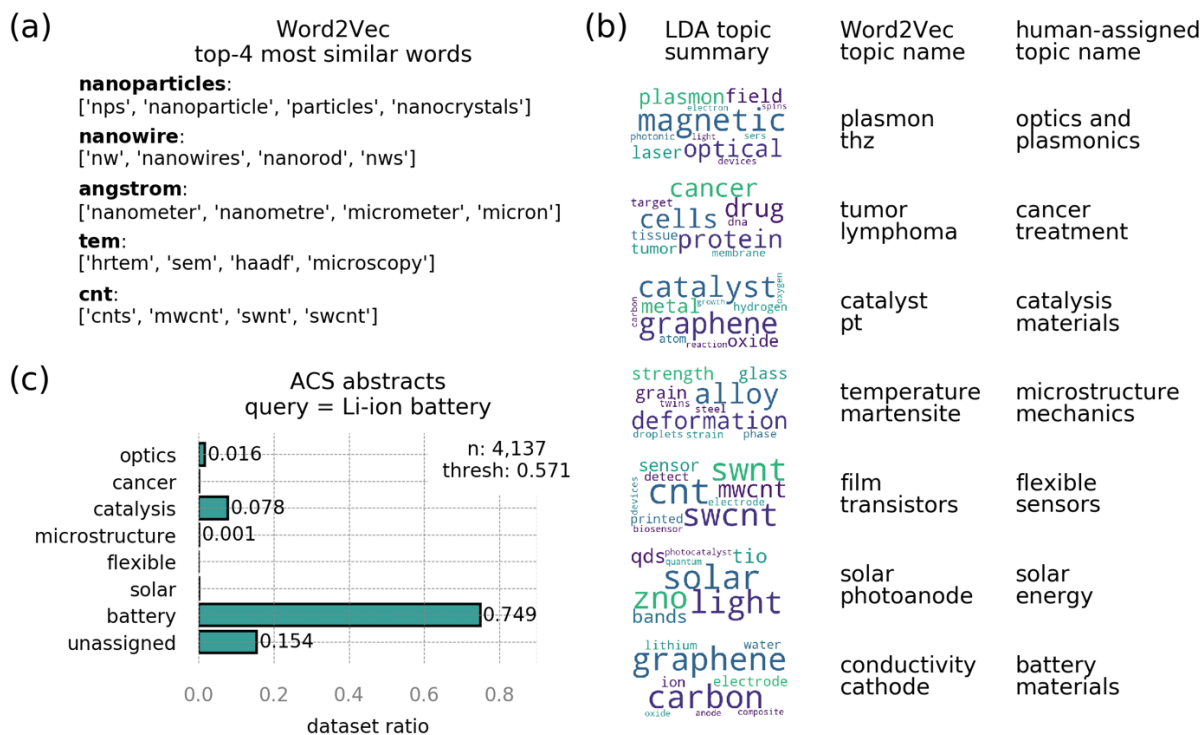

**Figure 6: Topic Modeling for Nanostructures Query.** (a) Word vectorization lookup examples for a Word2Vec model trained on the abstract and introduction texts from the nanostructure query. (b) LDA topic modeling applied to the introduction and abstract texts reveals some of the most popular technological applications of nanostructures. The Word2Vec topic name and human-assigned topic name represent the further attempt to summarize the words of the topics into a more concise two word title. (c) Distribution of topics assigned to a group of 4237 abstracts collected from a query of American Chemical Society (ACS) journals for Li-ion batteries.

*Topic Modeling.* When all of the abstract and introductory texts related to the general nanostructures *query* are collected together, certain topics (groups of related words) arise from both high-frequency word occurrences and common word orderings. Perhaps unsurprisingly, these topics reveal many of the popular technological applications of nanostructured materials, and when not explicitly included as part of the search *query* word families, can provide meaningful extra context. Figure 7b illustrates how Latent Dirichlet Analysis (LDA) [42], a popular technique used for topic modeling, is used in context of the nanostructure query. The word clouds, provided as a means to visualize the LDA output, illustrate the unsupervised clustering of related words into topics. Unfortunately, LDA does not provide a topic name to the words it clusters together. To this end, we illustrate how the trained Word2Vec model can be used to create topic names and how they closely mirror and/or are quickly resolved into rational human-suggested titles. For example, the topic containing "catalyst", "metal", "oxide", "graphene", "reaction", etc. is given the Word2Vec topic name of "catalyst", "pt", which is easily understood to represent the general class of "catalysis materials". To demonstrate that the LDA has indeed learned topics in some appropriate fashion, we collected 4137 abstract and introduction texts from a query of ACS journal family for "Li-ion batteries" and observed that the majority of the documents were categorized explicitly as belonging to batteries, or the highly related/overlapping catalysis category.

| extracted images | distributed captions | assigned labels |
|---|---|---|
| (a) 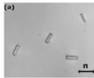 | (POM) images of silica rods of average length 6.5 μm and diameter 0.75 μm (designated as micro-rods) dispersed in a planar cell without crossed polarisers. | caption: 'silica' 'rods' 'planar' 'crossed' 'micro'<br>abstract: 'cell' 'elongated' 'cylindrical'<br>topic: 'optics and plasmonics' |
| (b) 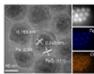 | HRTEM image of Fe/FeO NCs. Inset: EDS-mapping of Fe/FeO NCs. | caption: 'hrtem' 'eds' 'ncs' 'fe' 'mapping'<br>abstract: 'nanotherapeutics' 'icg' 'nanocarrier'<br>topic: 'cancer treatment' |
| (c) 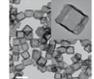 | TEM and magnified (inset) images of RuIrZnOx-U nanoboxes. | caption: 'tem' 'nanoboxes'<br>abstract: 'electrode' 'photocatalyst' 'air' 'synthetic'<br>topic: 'catalysis materials' |
| (d) 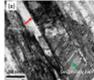 | The TEM bright field image is showing twinning blocky austenite grains after compression test at ~10−1/s (a). | caption: 'austenite' 'grains' 'bright' 'tem' 'compression'<br>abstract: 'twin' 'microstructure'<br>topic: 'microstructure mechanics' |
| (e) 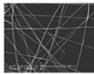 | The SEM images of AgNW thin film before electron beam irradiation | caption: 'beam' 'irradiation' 'agnw' 'film' 'sem'<br>abstract: 'substrate' 'nanowire'<br>topic: 'flexible sensors' |
| (f) 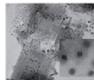 | TEM images of perovskite NCs. | caption: 'perovskite' 'ncs' 'tem'<br>abstract: 'mapbbr' 'photovoltaics'<br>topic: 'solar energy' |
| (g) 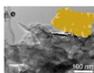 | The relationship between the initial structures of Si under a capacity restriction of 1500 mAh g−1. TEM images of aggregated lump (e). | caption: 'si' 'tem' 'capacity' 'structures' 'aggregated'<br>abstract: 'anodes'<br>topic: 'battery materials' |

**Figure 7: Example Image Labels after Hierarchical Label Assignment.** (a-g) Examples of hierarchical label assignment for images containing a properly distributed caption. For each image, the caption labels are limited to caption text only. Conversely, the abstract labels are free to draw additional relevant words from the abstract text, and the topic labels come from the human-assigned topic names from the LDA topic summaries. | DOIs for articles containing the example images (natural reading order): 10.1038/s41598-019-40198-1,10.1038/s41467-019-12142-4, 10.1038/s41467-019-12885-0, 10.1038/s41598-019-55803-6, 10.1038/srep17716, 10.1038/am2016167, 10.1038/srep42734.

*Hierarchical Label Assignment.* The task of resolving the sentence or phrases into a series of relevant image labels is handled using Word2Vec and LDA topic modeling at multiple structural levels. First, the explicit text of the caption must be processed to extract the most important nouns and adjectives that are indicative of the specific content of the given image. Figure 7 provides examples of how an image/caption pair is transformed into a series of hierarchical labels that describe and provide context for the image content. In all examples, the "caption" labels are determined using an iterative word dropout approach that removes words furthest away (measured by cosine similarity) from the center of the current group. The "abstract" labels are the 2-3 words closest to the center of the combination of abstract and caption words, and typically provide context for the image that is not found explicitly in the caption. Finally, the LDA model trained on the abstracts and introduction texts, assigns the best "topic" label to the document containing the image, if a confidence threshold of 0.80 is exceeded. Figure 7 overall provides several compelling examples of how this approach can provide useful contextual labels for the images from language outside that found explicitly in the distributed caption. For example, we learn things that we can confirm visually, such as the fact that nanorods are elongated and cylindrical (Figure 7a). We also learn things that an expert might know that are useful for understanding the function of the image content, such as the facts that: Fe/FeO nanocrystals function as nanotherapeutics (Figure 7b); RuIrZnO$_x$-U nanoboxes are synthetic as opposed to biological catalysts (Figure 7c); austenite grains describe the "microstructure" of the image (Figure 7d); nanocrystals are MA lead halide perovskite (numerical characters get stripped in the text preprocessing, so 'mapbbr' refers to for MAPbBr$_3$) (Figure 7f), and even in Figure 7e where the abstract context is more redundant than unique or complimentary, the topic provides useful context for where the specific image content appears from an application perspective.

There are a few subtle issues with some of the assigned labels that are a result of some of the known shortcomings in Word2Vec model training. Most notably is that in some cases, similarity is more indicative of how interchangeable/related words are, as opposed to measuring their actual likeness. For example, the nanoboxes in Figure 8c are inaccurately labeled as photocatalysts and should be described as *electro*catalysts. While these words are highly related and often found in interchangeable contexts (i.e. [photocatalysts/electrocatalysts] facilitate water-splitting … etc.), they can present certain instances where the suggested labeling in problematic. Overall, the combination of Word2Vec modeling with LDA topic discovery provides a solid backbone for self-labeling imaging effort. Future work will involve finding ways to use language components and the imaging analysis jointly to describe image content.

## Conclusion

We present EXSCLAIM!, a software pipeline for the automatic **EX**traction, **S**eparation, and **C**aption-based natural **L**anguage **A**nnotation of **IM**ages from scientific figures, detailing the specific extraction tools and providing quantitative measures of performance for image type classification and keyword labeling accuracy on a crowdsourced-labeled dataset, and an extracted dataset of nanostructure figures from Nature family journals, respectively. In addition, we provided discussions and useful model implementations aimed at assigning image labels from complete sentence text. Successful consolidation of images and self-labeling of images from scientific literature sources will not only enhance the navigation and searchability of images spanning materials, medical, and biological domains, but is a vital first step towards introducing scientific imaging to the canon of training datasets for state-of-the art deep learning and computer vision algorithms.

## Datasets and Code Availability

A dataset used to validate the classification and bounding box prediction accuracy of the FigureSeparator component of the EXSCLAIM! pipeline, as presented in Figure 3, can be found via the Materials Data Facility (https://doi.org/10.18126/a6jr-yfoq). A dataset illustrating how a sample query submitted to the EXSCLAIM! pipeline can be used to construct a sizable labeled dataset (> 280,000 images) of microscopy images from literature can be found via the Materials Data Facility (https://doi.org/10.18126/v7bl-lj1n). Analysis of image extraction and keyword relevance associated with this dataset is presented Figure 4 of the manuscript. All tools in the main extraction pipeline are available on github (https://github.com/MaterialEyes/exsclaim).

**Acknowledgement**

This material is based upon work supported by Laboratory Directed Research and Development (LDRD) funding from Argonne National Laboratory, provided by the Director, Office of Science, of the U.S. Department of Energy under Contract No. DE-AC02-06CH11357. Use of the Center for Nanoscale Materials, an Office of Science user facility, was supported by the U.S. Department of Energy, Office of Science, Office of Basic Energy Sciences, under Contract No. DE-AC02-06CH11357. We gratefully acknowledge the computing resources provided and operated by the Joint Laboratory for System Evaluation (JLSE) at Argonne National Laboratory. We gratefully acknowledge the computing resources provided on Bebop, a high-performance computing cluster operated by the Laboratory Computing Resource Center at ArgonneNational Laboratory.